\documentclass[twocolumn,showpacs,preprintnumbers,amsmath,amssymb]{revtex4}
\usepackage{cases} 

\begin{document}
\title{Quasirelativistic Langevin equation}
\author{A.V. Plyukhin}
\email{aplyukhin@anselm.edu}
 \affiliation{ Department of Mathematics,
Saint Anselm College, Manchester, New Hampshire 03102, USA 
}

\date{\today}% It is always \today, today,

\begin{abstract}
We address the problem of a microscopic derivation of the  
Langevin equation for a weakly relativistic Brownian particle. 
A non-covariant Hamiltonian model is adopted, in which
the free motion of particles is described relativistically, while their 
interaction  is treated classically, i.e., by means of 
action-to-a-distance interaction potentials. 
Relativistic corrections to the classical Langevin equation
emerge as nonlinear dissipation terms
and originate from the  nonlinear dependence of 
the relativistic velocity on momentum. On the other hand,
similar nonlinear dissipation forces also appear  as 
classical (non-relativistic)  
corrections to the weak-coupling approximation. 
It is shown that these classical corrections, which are usually ignored
in  phenomenological models, 
may be  of the same order of magnitude, if not larger than relativistic ones. 
The interplay of relativistic corrections and classical 
beyond-the-weak-coupling contributions
determines the sign of the leading nonlinear dissipation term in the Langevin
equation, and thus is qualitatively important.

%It is found that the interplay of relativistic and classical 
%nonlinear terms may cause the sign of a leading nonlinear 
%dissiaptive term to be temperature dependent.  

\end{abstract}


\maketitle
\section{Introduction}
Relativistic Brownian motion is the underpinning 
paradigm in several modern fields, 
including transport and thermalization  processes in  
quark-gluon plasma, astrophysical fluids, and graphene~\cite{Dunkel_review}.  
Despite a  high motivation
toward the  construction of a unifying approach,
there is currently no consensus on the form 
of   Langevin and master equations describing a relativistic Brownian 
particle. 
Several versions  
were proposed in recent years~\cite{Dunkel_review}, 
but their status and validity range are often obscure.
The difficulties are many, and some are fundamental 
to relativistic many-body dynamics~\cite{Dunkel_review,Hakim}.
In the nonrelativistic theory, the standard equations of Brownian motion 
can be derived microscopically, 
eliminating (fast) degrees of freedom of the thermal bath
with a projection operator or some other technique~\cite{Zwanzig}.
This is much harder to do in the relativistic domain because
the Lorentz-invariant dynamics of a system of particles 
also involves the degrees of freedom of the field 
through which the particles interact. 
For weakly relativistic systems
to second order in $v/c$, the elimination of field degrees of freedom 
is straightforward~\cite{Landau}, 
but comes at the expense of the emergence of additional  
velocity-dependent forces, which are difficult to handle for many-particle
systems within the Hamiltonian formalism~\cite{Orlov}.

Another dissonance with the standard classical approach comes from  the limited
validity of time-scale separation methods in the relativistic domain. 
For a non-relativistic Brownian particle  
the mean-square momentum is linear 
with  mass $\langle P^2\rangle\sim k_BTM$, 
which implies that at any given  temperature $T$ the 
average thermal momentum of the heavy Brownian particle is much larger 
(and the velocity is much smaller) than that of  particles of the bath with 
mass $m\ll M$.  This enables one to justify the weak-coupling approximation
to the lowest order in the small mass ratio parameter.
On the other hand, for  the ultra-relativistic particle with 
the thermal momentum much larger than $Mc$, 
the equipartition theorem [see Eq. (\ref{EQP1}) below] 
takes the form  that does not involve the mass of the particle
$c\langle |P|\rangle=k_BT$. 
Clearly, conventional time-scale separation methods 
cannot be applied in this case,  since heavy and light particles have 
comparable momenta. As will be shown below, a similar situation may take place 
also for a weakly relativistic Brownian particle when it is immersed in an
ultra-relativistic bath.

Despite  these difficulties 
(and  perhaps because of them),
many authors prefer to pursue an approach based on a straightforward
extension of the nonrelativistic Langevin phenomenology
~\cite{Deb,Zyg,quark,Deb2,Feld}.
In a simple version, one assumes that the dissipative force on the
particle is linear in the particle's velocity $V$ and composes the Langevin
equation for the particle's momentum $P$ in the rest frame of the bath
in the form~\cite{Deb}
\begin{eqnarray}
\frac{dP}{dt}= -\zeta \, V(P) +\xi(t),
\label{LE0}
\end{eqnarray}
where $\xi(t)$ is a stationary zero-centered 
delta-correlated (white) noise, 
\begin{eqnarray} 
\langle \xi(t)\xi(t')\rangle=2D\,\delta(t-t').
\label{delta}
\end{eqnarray}
For the relativistic domain, Eq. (\ref{LE0}) is nonlinear since
velocity is a nonlinear function of momentum,  
\begin{eqnarray}
V(P)=\frac{dE}{dP}=\frac{c^2 P}{E}=\frac{1}{\Gamma(P)}\,\frac{P}{M},
\label{velocity}
\end{eqnarray}
where $E$ is the energy of a free particle
\begin{eqnarray}
E(P)=\sqrt{c^2P^2+M^2c^4}=Mc^2\,\Gamma(P),
\label{energy}
\end{eqnarray}
and
\begin{eqnarray}
\Gamma(P)=\sqrt{1+\left(\frac{P}{Mc}\right)^2}.
\label{Gamma}
\end{eqnarray}
Since   Eq. (\ref{LE0}) is not amenable to closed-form
analytic solutions, 
a fluctuation-dissipation relation  between 
the friction coefficient $\zeta$
and the strength of noise $D$  in general cannot be established.  
However, further progress can be achieved under two additional assumptions.
The first one is that 
the random force $\xi(t)$ is a Gaussian process, i.e., vectors 
of observed values $\{\xi(t_1), \dots, \xi(t_n)\}$ have a multivariate
normal distribution. 
As known from the general theory~\cite{Zwanzig}, 
in this case the 
corresponding Fokker-Planck equation for the distribution function $f(P,t)$
has the form 
\begin{eqnarray}
\frac{\partial}{\partial t}f(P,t)=
\zeta\, \frac{\partial}{\partial P}\left\{
V(P)\,f(P,t)
\right\}
+D\,\frac{\partial^2}{\partial P^2}f(P,t)
\label{FPE}
\end{eqnarray}
for any function $V(P)$, linear or not.
The second assumption is that
the stationary solution of this equation
$f(P)=C\, \exp\left[-\frac{\zeta}{D}\,E(P)\right]$
must coincide with the 
Maxwell-J\"uttner distribution
\begin{eqnarray}
\rho_{MJ}(P)=Z^{-1}\,e^{-\beta \,E(P)},
\label{MJ}
\end{eqnarray}
where $\beta=1/k_BT$ is the inverse temperature of the bath 
in the bath's reference frame, and $E(P)$ is given by (\ref{energy}).
This immediately gives the
fluctuation-dissipation relation  
\begin{eqnarray}
\zeta=\beta \,D.
\label{FDT0}
\end{eqnarray}

While attractively simple, 
the above phenomenological scheme suggests no clue about its 
range of validity. 
It also appears to be unnecessarily restrictive in its demand of 
the noise to be Gaussian. Within a nonrelativistic theory, both 
phenomenological
and  microscopic, the assumption of Gaussian noise is
unnecessary to  derive the fluctuation-dissipation relation.
It is therefore natural to ask if, and under what conditions,  
the Langevin (\ref{LE0}) and 
Fokker-Planck (\ref{FPE}) equations
can be derived microscopically. 
As mentioned above, severe difficulties of the relativistic
theory of many-body interacting systems   generally make such a derivation
hardly possible.  
However, one may expect that some difficulties
can be avoided  for systems with contact interactions, i.e. when
particles interact via point-like binary 
collisions~\cite{Dunkel_paper1,Dunkel_paper2}. 
In this case,
interactions can be fully described by conservations laws, and 
one can avoid the infamous problem of constructing a relativistic
action-at-a-distance Hamiltonian of many interacting  particles.

Following this line, Dunkel and H\"anggi~\cite{Dunkel_paper2} 
discussed the derivation of the relativistic Langevin equation 
for the Rayleigh model, in which a Brownian particle 
interacts with bath molecules via elastic instantaneous 
(and therefore binary) collisions. 
As is well known, for the classical Rayleigh model the noise is not 
Gaussian~\cite{Kampen_paper,Plyukhin_FP}. 
The derivation presented in~\cite{Dunkel_paper2} emphasizes 
the non-Gaussian nature of the noise for the relativistic domain.
The authors  employed a 
nonperturbative approach that leads to a rather complicated expression
for the dissipating force $F_{diss}$, which is 
amenable  only to numerical evaluation.
This makes it difficult to verify the validity
of the phenomenological ansatz 
$F_{diss}=-\zeta\, V(P)$ 
and the fluctuation-dissipation relation (\ref{FDT0}).

In this paper we address the problem of a microscopic derivation of 
the relativistic Langevin  from different premises. 
Namely, we consider 
a Hamiltonian 
model in which only the free motion of particles is treated relativistically, 
while the interaction is described classically, i.e.,  by means of 
action-at-a-distance potentials. Such  an approximation, 
which we  refer to as
quasirelativistic,  was  recently discussed and tested in~\cite{Morriss}. 
It produces, of course, noncovariant equations of motion, yet may be
acceptable for systems with very-low-density and/or short-range 
interactions. Numerical simulation shows 
that quasirelativistic many-particle system equilibrates
toward the Maxwell-J\"uttner distribution~\cite{Morriss}, 
which is similar to a fully relativistic molecular dynamics 
simulation~\cite{MJ_simulation}. 
Intuitively, 
in the limit when the range of interaction goes to zero,
one can expect to get the same results as for relativistically consistent 
models with
contact interaction. The advantage of the quasi-relativistic approach is 
that it enables one to apply well-developed perturbation techniques
of Hamiltonian theory of non-relativistic 
Brownian motion~\cite{MO,Albers}.  These methods
are not easy  to use
within models with instantaneous binary
collisions~\cite{Dunkel_paper1,Dunkel_paper2}  due to the presence 
of singular $\delta$-like forces.  
 
We shall assume that the rest mass $M$ of a Brownian particle is much larger
than the mass $m$  of a bath particle, so that the mass ratio parameter
$\lambda$ is small, 
\begin{eqnarray}
\lambda=\sqrt{\frac{m}{M}}\ll 1.
\label{lambda}
\end{eqnarray}
Also we shall restrict the discussion to temperature regimes for which the
characteristic thermal momentum of the Brownian particle $P_T$ is much larger
than that of a bath particle $p_T$ and much smaller than $Mc$,
\begin{eqnarray}
p_T\ll P_T\ll  Mc. 
\label{order}
\end{eqnarray}
This will allow us to construct a perturbation technique
similar to that for the non-relativistic theory.  
As will be shown, the condition (\ref{order}) is not too restrictive:
While the Brownian particle is assumed to be weakly relativistic, 
%$P_T\ll Mc$, 
particles of the bath may be weakly, moderately, or 
even ultra relativistic.

We shall show that under the above assumptions
the phenomenological Langevin equation (\ref{LE0}) and the
fluctuation-dissipation relation (\ref{FDT0}) are not valid
for any regime for which nonlinearity of the function $V(P)$
is essential. The comparison of phenomenological and microscopic
predictions is easier if, given the condition (\ref{order}), 
one retains only the leading nonlinear term 
in the expansion of $V(P)$, 
\begin{eqnarray}
V(P)=\frac{1}{\Gamma(P)}\, \frac{P}{M}\approx\left[
1-\frac{1}{2}\, \left(\frac{P}{Mc}\right)^2
\right]\, \frac{P}{M}.
\label{Vexp}
\end{eqnarray}
With the approximation (\ref{Vexp}) and relation (\ref{FDT0}), 
the phenomenological Langevin equation
(\ref{LE0}) takes the form
\begin{eqnarray}
\frac{d}{dt}P(t)=-\gamma_1\, P(t)-\gamma_2 \,P^3(t) +\xi(t),
\label {LE1}
\end{eqnarray}
with damping coefficients
\begin{eqnarray}
\gamma_1=\frac{\beta\, D}{M},\quad\quad\quad
\gamma_2=-\frac{\beta\, D}{2\,M^3\,c^2}<0.
\label{gamma12}
\end{eqnarray}
%The ratio $\gamma_1/\gamma_2=-2M^2c^2$
%is negative and temperature independent. 
The microscopic theory developed below
also leads to the Langevin equation in the form (\ref{LE1}), but 
with fluctuation-dissipation relations different and more complicated 
than (\ref{gamma12}).
Note that 
the nonlinear term in Eq.(\ref{LE1}) 
originates from the first relativistic
correction to the classical linear relation $V=P/M$.
% i.e. from the second term in(\ref{Vexp}).  
On the other hand, 
from the microscopic theory of non-relativistic 
Brownian motion it is known that similar 
nonlinear dissipation terms also appear
in the Langevin equation beyond the weak-coupling limit.
These contributions, 
which are missing in phenomenological Langevin 
equations (\ref{LE0}) and (\ref{LE1}), 
are of classical nature and 
originate from higher-order terms in the expansion of the particle's 
propagator  in powers of the mass ratio parameter $\lambda$. 
We shall show that these classical nonlinear corrections 
are of the same order of magnitude or larger than the  
corresponding relativistic contributions. 
A consistent theory, which  takes into account the
interplay of both relativistic and classical contributions for the  nonlinear
dissipative force $F_{diss}$, does not support 
the simple ansatz $F_{diss}\sim V(P)$
adopted in the phenomenological theory.

One  prediction of the presented theory is that the sign
of the nonlinear damping coefficient $\gamma_2$ in Eq. (\ref{LE1})
is not predetermined and may depend on temperature
and a detailed form of the microscopic correlations. This is in contrast to
the second of the phenomenological relations (\ref{gamma12}), which predicts
that $\gamma_2$ is negative.
For $\gamma_2<0$, 
one can show that the Langevin equation (\ref{LE1}), as well as 
the corresponding
Fokker-Planck equation, leads to an 
ill-behaved  stationary distribution $f(P)$ diverging for large $P$.  
Thus, in the phenomenological theory, the approximation (\ref{Vexp}) is
insufficient and one needs to 
retain nonlinear terms of higher orders in $P/Mc$.
In contrast, the presented microscopic 
theory predicts that
for certain temperature intervals $\gamma_2$ may be positive 
and the Langevin equation (\ref{LE1})
has meaningful equilibrium properties.
Other implications are discussed in the last Sec. VIII.

\section{Scaling relations}
We consider a Brownian particle 
(below referred to for short 
as {\it the} particle) that  is not too far from the equilibrium 
in which the momentum
distribution is given by the Maxwell-J\"uttner distribution (\ref{MJ}). 
A relativistic version of 
the equipartition theorem for the particle in equilibrium in one dimension has 
the form
\begin{eqnarray}
\langle V(P)\, P\rangle=
\left\langle \frac{1}{\Gamma(P)}\,\frac{P^2}{M}\right\rangle =
\frac{1}{\beta},
\label{EQP1}
\end{eqnarray}
where $\Gamma(P)$ is given by (\ref{Gamma})
and the angular brackets mean the average with the distribution (\ref{MJ}).
Unlike its classical counterpart (when $\Gamma\to 1$), 
the relativistic equipartition
relation (\ref{EQP1}) does not allow one to find  an exact expression for 
the thermal momentum of the particle 
$P_T=\sqrt{\langle P^2\rangle}$.
Yet Eq. (\ref{EQP1})  is convenient to evaluate 
an approximate value of $P_T$ as follows.
Let us define the parameters
\begin{eqnarray}
\epsilon=\sqrt{\frac{1}{\beta \,m\,c^2}},\qquad
\delta=\lambda\,\epsilon =\sqrt{\frac{1}{\beta \,M\,c^2}}, 
\label{epsilondelta}
\end{eqnarray}
characterizing the strength of
relativistic effects for the bath and the particle, respectively.
Using approximations 
$\Gamma\sim 1$  for  $\delta \lesssim 1$ 
and  $\Gamma(P)\sim P/Mc$  for $\delta \gg 1$, from (\ref{EQP1}) one obtains
\begin{subnumcases} 
{P_T\approx}
\sqrt{\frac{M}{\beta}}=\delta\, Mc,& for  $\delta \lesssim 1$,\\
\frac{1}{c\,\beta}=\delta^2\, Mc, & for $\delta \gg 1$.
\end{subnumcases}

The validity of this estimation can be verified 
by direct evaluation of the mean-square momentum for 
the Maxwell-J\"uttner
equilibrium
\begin{eqnarray}
\langle P^2\rangle=Z^{-1}\int e^{-\beta E(P)}P^2\,dP.
\label{aux00}
\end{eqnarray}
Indeed, in one dimension from (\ref{aux00}) one obtains exactly
\begin{subnumcases} 
{P_T=\sqrt{\langle P^2\rangle}=}
\sqrt{\frac{M}{\beta}}\,\,\phi_1(\delta),& or\\
\frac{1}{c\,\beta}\,\,\phi_2(\delta), & 
\end{subnumcases}
with dimensionless functions
\begin{eqnarray}
\phi_1(\delta)=\left[
\frac{K_2(1/\delta^{2})}{K_1(1/\delta^{2})}
\right]^{1/2},
\qquad
\phi_2(\delta)=\frac{1}{\delta}\,\phi_1(\delta).
\end{eqnarray}
Here  $K_n(x)$ are the modified Bessel functions of the second kind.
As can be checked,  $\phi_1(\delta)\sim 1$ for $\delta\lesssim 1$ and  
$\phi_2(\delta)\sim 1$ for $\delta\gg 1$, so that the exact relations 
(18) lead to  the estimations (16).
A similar consideration can be carried out 
to evaluate  the 
thermal momentum $p_T=\sqrt{\langle p^2\rangle}$
of a bath particle 
\begin{subnumcases} 
{p_T\approx}
\sqrt{\frac{m}{\beta}}=\epsilon\, mc,& for  $\epsilon \lesssim 1$,\\
\frac{1}{c\,\beta}=\epsilon^2\, mc, & for $\epsilon \gg 1$.
\end{subnumcases}

In order to design an appropriate perturbation technique, we need to
establish relations between $P_T$ and $p_T$ for different temperature regimes.
We shall use the following nomenclature.

\vspace{2mm}

{\it{Regime A}}
is defined by relation
\begin{eqnarray}
\epsilon\ll 1,
\label{A1}
\end{eqnarray}
or $k_BT\ll mc^2$. 
Since the other relevant 
parameter is also small $\delta=\lambda\,\epsilon\ll 1$, 
in this regime   relativistic effects are weak for both the bath and 
the particle. As follows from (16) and (20),  
the thermal momentum of a bath particle is
$\lambda$ times smaller that of the  particle,
\begin{eqnarray}
p_T=\lambda\, P _T.
\label{scaleAB}
\end{eqnarray} 

\vspace{2mm}

{\it{Regime B}} \, is defined by the condition
\begin{eqnarray}
\epsilon\sim 1,
\label{B}
\end{eqnarray}
or $k_BT\sim mc^2$.   
The other relevant parameter $\delta$ is small
$\delta=\lambda\,\epsilon\sim\lambda\ll 1$. This regime corresponds to the
moderately relativistic bath and weakly relativistic particle.  
The relation between $p_T$ and $P_T$ is still given by (\ref{scaleAB}).
It is therefore convenient for both regimes  $A$ and $B$
to introduce the particle's scaled momentum 
\begin{eqnarray}
P_*=\lambda\, P,
\label{PA}
\end{eqnarray}
which on average is  expected to be of the same order of magnitude 
as the thermal momentum of a bath particle.

%Dispite of the similarity, 
%the perturbation analysis of regimes $A$ and $B$ is slightly different,
%so it is convenient to consider the two regimes seperately. 

\vspace{2mm}

{\it{Regime  C}} \, corresponds to a sub-domain
of the ultra-relativistic bath  defined by the relation
\begin{eqnarray}
1\ll\epsilon\ll \lambda^{-1},
\label{C}
\end{eqnarray}
or  $mc^2\ll k_BT\ll Mc^2$.
The right-hand side of the inequality (\ref{C}) 
ensures that $\delta\ll 1$, so that 
the particle is still weakly relativistic and its 
thermal momentum $P_T$ is given by  the classical expression (16a).  
In contrast, since $\epsilon\gg 1$, 
the thermal momentum of a bath particle
is given by the ultrarelativistic expression (20b). Here 
$p_T$ is still smaller than
$P_T$ but now with the scaling factor $\delta$, 
\begin{eqnarray}
p_T=\delta \cdot P_T.
\label{scaleC}
\end{eqnarray}
For this regime we define 
the scaled momentum of the particle as
\begin{eqnarray}
P_*=\delta\cdot P
\label{PCD}
\end{eqnarray}
with the expectation that on average $P_*$ is of the same order of magnitude as
momenta of bath particles.

\vspace{2mm}

{\it{Regime  D}} \, is defined by the relation
\begin{eqnarray}
\epsilon\sim \lambda^{-1},
\label{D}
\end{eqnarray}
or $k_BT\sim Mc^2$. In this case $\epsilon\gg 1$ and $\delta\sim 1$,
which corresponds to the ultra-relativistic bath and the moderately 
relativistic particle. Thermal momenta $p_T$ and $P_T$ are the same as for
regime $C$, related as $p_T=\delta \cdot P_T$ and, since $\delta\sim 1$,
are of the same order of magnitude.  

\vspace{2mm}

{\it{Regime  E}} \, is defined by
\begin{eqnarray}
\epsilon\gg \lambda^{-1}.
\label{E}
\end{eqnarray}
Since $\epsilon, \delta\gg1$, for this regime
both the particle and bath are ultra-relativistic.
Thermal momenta of the particle
and of the bath are given by ultra-relativistic expressions (16b) and 
(20b), respectively, and as for regime $D$  
are of the same order of magnitude.

In the next section we use the above scaling relations to formulate 
quasi-relativistic dynamic equations in a form that explicitly involves
a small parameter relevant to a given temperature regime.
We shall restrict ourselves to regimes $A$, $B$, and $C$ only,
for which $P_T\gg p_T$.
Regimes $D$ and $E$, for which $P_T\sim  p_T$,  
cannot be treated with the conventional perturbation techniques
and will not be discussed further.

\section{Quasi-relativistic Hamiltonian}
Let $(X,P)$ and $\{x_i,p_i\}$ be the sets of coordinates and momenta of
the particle  
and particles of the thermal bath, respectively.
The motion will be assumed to occur in one spatial dimension, but  
this assumption is not essential and is adopted merely to simplify notations.
The quasi-relativistic Hamiltonian~\cite{Morriss} 
of the combine system of the particle and the bath is
\begin{eqnarray}
H=E(P)+H_0,
\label{H}
\end{eqnarray}
where $E(P)$ is the energy of the free particle given by (\ref{energy}), 
and $H_0$ is the Hamiltonian of the bath interacting  
with the particle fixed at the position $X$,
\begin{eqnarray}
H_0=\!\sum_{i}e(p_i)+U(X).
\label{hamiltonian_0}
\end{eqnarray}
In this expression $e(p_i)$  is the energy of $i$-th free particle of the bath,
\begin{eqnarray}
e(p_i)=\sqrt{c^2p_i^2+m^2c^4},
\end{eqnarray}
and the potential $U(X)=U(X,\{x_i\})$ 
describes  the  interaction of the particle with  the bath, as well as  
bath particles with each other. 
The interaction is  
understood classically as action at a distance,
no Darwin-like momentum-dependent corrections~\cite{Landau,Orlov} are included 
in the potential $U$.
Thus the only difference between our quasirelativistic Hamiltonian 
and that of the nonrelativistic theory is a nonquadratic
dependence of free particle energy  terms $E(P)$ and $e(p_i)$ on momenta. 
Bath particles will be  assumed to have the same rest mass $m$, which is 
much smaller than that of the particle $M$, so that 
$\lambda=\sqrt{m/M}\ll 1$.

The Liouville operator corresponding to the Hamiltonian (\ref{H}) splits
naturally in two parts
\begin{eqnarray}
L=L_0+L_{part}.
\label{Ltotal}
\end{eqnarray}
The Liouville  operator $L_0$ governs  the dynamics of the bath with Hamiltonian
$H_0$, 
\begin{eqnarray}
L_0=\sum_{i}v_i\frac{\partial}{\partial x_i}+f_i\frac{\partial}{\partial
  p_i}.
\label{Lbath}
\end{eqnarray}
Here $f_i=-\partial H_0/\partial x_i$ is the force on a bath particle
and a bath particle velocity as a function of momentum is
\begin{eqnarray}
v_i=\frac{\partial H_0}{\partial p_i}=
\frac{1}{\gamma (p_i)}\,\frac{p_i}{m}
\end{eqnarray}
with 
\begin{eqnarray}
\gamma (p)=\sqrt{1+\left(\frac{p}{m\,c}\right)^2}.
\end{eqnarray}
The operator $L_{part}$ involves derivatives with respect to the coordinate 
and momentum of the particle
\begin{eqnarray}
L_{part}= V(P)\,\frac{\partial}{\partial X}+F\frac{\partial}{\partial P}. 
\label{Lpart}
\end{eqnarray}
Here $F=-\partial H/\partial X$ is the forces on the particle, and  
\begin{eqnarray}
V(P)=\frac{\partial H}{\partial P}=\frac{1}{\Gamma(P)}\,\,\frac{P}{M},
\label{V}
\end{eqnarray}
with $\Gamma(P)$ given by (\ref{Gamma}), 
is the particle's velocity.

The only difference between the quasirelativistic Liouville
operator $L$ 
and its nonrelativistic counterpart is the presence in the above
formula of dimensionless factors $\Gamma^{-1}(P)$ and $\gamma^{-1}(p)$, 
which makes velocities nonlinear
functions of momenta.

The next step is to write the Liouville operator $L$ in terms of the scaled
momentum of the particle $P_*$, which would put  $L$  into  a form 
that explicitly involves a relevant 
small parameter.

\vspace{0.2cm}  
 
{\it\underline {In regimes  A and B}} ($\epsilon\lesssim 1$), 
since $\delta=\lambda\,\epsilon\ll 1$,  the particle 
is weakly relativistic and
\begin{eqnarray}
P_T=\sqrt{M/\beta}=\delta\cdot Mc\ll Mc,
\end{eqnarray}
[see Eq. (16a)].  
Then one can use the approximation
\begin{eqnarray}
\frac{1}{\Gamma(P)}\approx 
1-\frac{1}{2}\,\left(\frac{P}{Mc}\right)^2,
\label{expansionA}
\end{eqnarray} 
which also can be written as
\begin{eqnarray}
\frac{1}{\Gamma(P)}\approx 
1-\frac{\delta^2}{2}\cdot
\left(
\frac{P_*}{p_T}
\right)^2.
\label{expansionA2}
\end{eqnarray}
As discussed in
the previous section,
for these regimes the scaled momentum of the particle $P_*$
and the bath's thermal momentum are defined as
\begin{eqnarray}
P_*=\lambda P,\qquad p_T=\sqrt{\frac{m}{\beta}}=\epsilon m c.
\end{eqnarray}
With the approximation (\ref{expansionA2}), 
the particle's velocity reads
\begin{eqnarray}
V=\frac{1}{\Gamma(P)}\,\frac{P}{M}\approx
\lambda\left\{
 1-\frac{\delta^2}{2}\,\left(\frac{P_*}{p_T}\right)^2
\right\}\, \frac{P_*}{m},
\label{Vapprox}
\end{eqnarray}
and the operator $L_{part}$ [ Eq.(\ref{Lpart})] takes the form 
\begin{eqnarray}
L_{part}=\lambda\, L_1+\lambda\,\delta^2\, L_2,
\end{eqnarray} 
with the classical part
\begin{eqnarray}
L_1= 
\frac{P_*}{m}\,\frac{\partial}{\partial X}+F\,\frac{\partial}{\partial P_*} 
\label{L1A}
\end{eqnarray}
and the relativistic correction
\begin{eqnarray}
L_2= -\frac{1}{2\,m\,p_T^2}\,\,P_*^3\,\,\frac{\partial}{\partial X}. 
\label{L2A}
\end{eqnarray}
Thus, for regimes $A$ and $B$ the Liouville operator $L$ for the total system
(\ref{Ltotal}) 
can be written as
\begin{eqnarray}
L=L_0+\lambda\, L_1+\lambda\,\delta^2\, L_2
\label{LAB}
\end{eqnarray}
with  $L_0, L_1$, and $L_2$  defined by (\ref{Lbath}),
(\ref{L1A}) and (\ref{L2A}), respectively. 

\vspace{0.2cm}

{\it\underline {In regime  C }} ($1\ll\epsilon\ll \lambda^{-1}$), 
since $\delta=\lambda\,\epsilon\ll 1$, 
the particle is still weakly relativistic $P_T\ll Mc$ 
and the approximation (\ref{expansionA}) for $\Gamma^{-1}$
is meaningful.
One can check that the expression (\ref{expansionA2})  
retains its form, although  now 
 the scaled momentum of the particle $P_*$
and the bath's thermal momentum are defined as
\begin{eqnarray}
P_*=\delta \cdot P,\qquad p_T=\frac{1}{c\,\beta}=\epsilon^2 m c,
\end{eqnarray}
as prescribed by Eqs. (\ref{scaleC}) and (\ref{PCD}) in the previous section.
The particle velocity now has the form
\begin{eqnarray}
V=\frac{1}{\Gamma}\,\frac{P}{M}\approx
\delta\left\{
 1-\frac{\delta^2}{2}\,\left(\frac{P_*}{p_T}\right)^2
\right\}\, \frac{P_*}{p_T}\, c,
\label{Vapprox2}
\end{eqnarray}
and the operator $L_{part}$ [Eq. (\ref{Lpart})] reads
\begin{eqnarray}
L_{part}=\delta\cdot L_1+\delta^3\cdot L_2
\end{eqnarray} 
where
\begin{eqnarray}
L_1= c\,\frac{P_*}{p_T}\,\frac{\partial}{\partial X}
+F\,\frac{\partial}{\partial P_*}, 
\label{L1C}
\end{eqnarray}
and
\begin{eqnarray}
L_2= -\frac{c}{2}\,\,\left(\frac{P_*}{p_T}\right)^3\,\,
\frac{\partial}{\partial X}. 
\label{L2C}
\end{eqnarray}
Thus, for regime $C$ the Liouville operator $L$ for the whole system
(\ref{Ltotal}) 
takes the form
\begin{eqnarray}
L=L_0+\delta\cdot L_1+\delta^3\cdot L_2,
\label{LC}
\end{eqnarray} 
where $L_0$, $L_1$, and $L_2$ 
are given by (\ref{Lbath}),
(\ref{L1C}), and (\ref{L2C}), respectively.

It is worthwhile to observe that the above relations
can be obtained from the corresponding expressions
for regimes $A$ and $B$ by making the replacements
\begin{eqnarray}
\lambda\to\delta, \quad m\,\to \,\frac{p_T}{c}.
\label{transition}
\end{eqnarray}

In Eqs. (\ref{LAB}) and (\ref{LC}), the dependence of the Liouville
operator $L$ on small parameters is explicit, 
which makes these expressions convenient for developing a 
perturbation technique, as discussed in the following sections.

\section{Pre-Langevin equation}
 In this section we apply the Mazur-Oppenheim projection operator 
technique~\cite{MO} to modify the exact equation of motion for the scaled 
momentum of the particle $P_*$ into a  form convenient for 
the subsequent derivation of the Langevin equation with a perturbation method.
%The procedure is well-known and popular, yet we shall briefly recapitulate
%the derivation.

{\it\underline {In regimes A \& B.}}  
$P_*=\lambda\, P$ and the equation of motion is   
\begin{eqnarray}
\frac{d}{dt} P_*(t)=\lambda\, F(t)=\lambda\, e^{Lt}F,
\label{eqofmotion1}
\end{eqnarray}
where $L$ is given by (\ref{LAB}) and  $F=F(t=0)$.
The propagator $e^{Lt}$ can be decomposed  as 
 \begin{eqnarray}
e^{Lt}=
e^{{\mathcal Q} L t}+\int_0^t e^{L(t-\tau)}\,\mathcal P L\, 
e^{{\mathcal Q} L \tau}\, d\tau,
\label{aux1}
\end{eqnarray}   
where $\mathcal Q=1-\mathcal P$ and $\mathcal P$ is an arbitrary operator.
This follows from 
the operator identity
\begin{eqnarray}
e^{(\mathcal A+\mathcal B)t}=e^{\mathcal At}+\int_0^t e^{\mathcal A (t-\tau)}
\mathcal B e^{(\mathcal A+\mathcal B )\tau}\, d\tau,
\label{aux2}
\end{eqnarray} 
with $A=L$, and $B=-\mathcal P L$. 
Inserting (\ref{aux1}) into (\ref{eqofmotion1}) yields
\begin{eqnarray}
\frac{d}{dt}P_*(t)=\lambda F^{\dagger}(t)+
\lambda \int_0^t e^{L(t-\tau)} \mathcal P L F^{\dagger}(\tau)\, d\tau,
\label{aux3}
\end{eqnarray}
where the projected force  is
\begin{eqnarray}
F^{\dagger}(t)=e^{\mathcal Q L t}F.
\end{eqnarray}

%Note that since $\mathcal{
%  P Q}=0$ and $\langle F\rangle=0$, the force $F^{\dagger}(t)$ is 
%zero centered,
%$\langle F^{\dagger}(t)\rangle=\mathcal P F^{\dagger}(t)=0$.

We shall assume that the initial distribution for bath
degrees of freedom is
\begin{eqnarray}
\rho_0=Z^{-1}\, e^{-\beta H_0}.
\end{eqnarray}
Hereafter the angular brackets  will denote
the average with the distribution $\rho_0$.
We  define the operator $\mathcal P$ to be  
the projection operator ($\mathcal P^2=\mathcal P$) that averages
over the initial degrees of freedom of the bath
\begin{eqnarray}
\mathcal P \,(\cdots)=\int \rho_0 \,(\cdots) 
\, \prod_i dx_idp_i=\langle \cdots\rangle. 
\end{eqnarray} 
The major benefit of this choice for $\mathcal P$ is the orthogonality 
relation
\begin{eqnarray}
\mathcal P L_0=0, 
\end{eqnarray}
which makes in the equation of motion (\ref{aux3}) 
a crucial reduction:
\begin{eqnarray}
\frac{d}{dt}P_*(t)
&=&\lambda \, F^{\dagger}(t)
\label{aux4}
\\
&+&
\lambda^2\int_0^t
e^{L(t-\tau)} \mathcal P\, (L_1+\delta^2 L_2)\, F^{\dagger}(\tau)\, d\tau.
\nonumber
\end{eqnarray}
Now the integral term in this equation does not 
involve derivatives with respect to the  
bath degrees of freedom (except for in the propagator $e^{Lt}$).

The next step is to take into account the explicit expressions
for $L_1$ and $L_2$ given by (\ref{L1A}) and (\ref{L2A})
and also the relation
\begin{eqnarray}
\mathcal P \frac{\partial}{\partial X} (\dots)=
\left\langle\frac{\partial}{\partial X} \,(\dots)\right\rangle=
-\beta\,\langle F \dots \rangle, 
\end{eqnarray}
which can be proved by integration by parts. This 
puts Eq.(\ref{aux4})
into the ``pre-Langevin'' form
\begin{eqnarray}
&&\frac{d}{dt}P_*(t)=\lambda\,F^{\dagger}(t)+
\lambda^2\int_0^t d\tau\,e^{L(t-\tau)}
\label{exacteqAB}\\
&&\times
\left\{
-\frac{1}{p_T^2}P_*+\frac{\partial}{\partial P_*}
+\delta^2\,\frac{1}{2p_T^4}\,P_*^3
\right\}\,
\langle F F^{\dagger}(t) \rangle.\nonumber
\end{eqnarray} 
Recall that for the given regimes $p_T=\sqrt{m/\beta}$.
Since $\mathcal{ P Q}=0$ and $\langle F\rangle=0$, 
the projected force $F^{\dagger}(t)$ is zero centered,
$\langle F^{\dagger}(t)\rangle=\mathcal P e^{\mathcal Q LT}F=0$.

{\it\underline {In regime C}} the scaled momentum is
$P_*=\delta\cdot P$, and the equation of motion is
\begin{eqnarray}
\frac{d}{dt} P_*(t)=\delta\cdot F(t)=\delta\cdot e^{Lt}F
\end{eqnarray} 
where $L$ is now given by (\ref{LC}). Then the  
same procedure as the one described above for regimes $A$ and $B$ 
puts the equation of motion
into the  form  
\begin{eqnarray}
&&\frac{d}{dt}P_*(t)=\delta\cdot F^{\dagger}(t)+
\delta^2\cdot\int_0^t d\tau\,e^{L(t-\tau)}
\label{exacteqC}\\
&&\times
\left\{
-\frac{1}{p_T^2}P_*+\frac{\partial}{\partial P_*}
+\delta^2\cdot\frac{1}{2p_T^4}\,P_*^3
\right\}\,
\langle F F^{\dagger}(t) \rangle.\nonumber
\end{eqnarray} 
This equation is similar to  Eq.(\ref{exacteqAB})
for regimes $A$ and $B$ except that
$\lambda$ is now replaced by $\delta$ and  
the thermal momentum of a bath particle is 
$p_T=1/c\,\beta$ instead of $p_T=\sqrt{m/\beta}$ for
regimes $A$ and $B$.

The only approximation made so far is the truncated expansion
(\ref{expansionA}) of $\Gamma^{-1}(P)$ for a weakly
relativistic particle. Otherwise,
the equations of motion (\ref{exacteqAB}) and (\ref{exacteqC})
are exact. Compared to the corresponding non-relativistic equations,
they contain an additional nonlinear term cubic in $P_*$. 
In order to make further progress and to put 
these equations into the Langevin 
form one needs to  expand $F^{\dagger}(t)$ in powers
of a relevant small parameter. As can be observed from  
(\ref{exacteqAB}) and (\ref{exacteqC}), 
higher-order terms of 
this expansion must be taken into account in order to consistently 
retain  the leading nonlinear relativistic correction.

\section{Langevin equation: regimes $A$ \& $B$}            
Let us find the perturbation expansion  
of the projected
force $F^{\dagger}(t)$  for regimes $A$ and $B$, when the Liouville operator
$L$ is given by (\ref{LAB}),
$L=L_0+\lambda\,L_1+\lambda\,\delta^2\,L_2$. Since $\mathcal P L_0=0$ and
$\mathcal QL_0=L_0$, we can write
\begin{eqnarray}
F^{\dagger}(t)=e^{\mathcal Q L t}F=e^{(L_0+\lambda \mathcal Q L_1
+\lambda\delta^2 \mathcal Q  L_2)\,t}F.
\end{eqnarray}
Next, 
as follows from the operator identity (\ref{aux2}), the part of the propagator 
involving $L_2$ gives a contribution 
of order $\lambda\,\delta^2$,
\begin{eqnarray}
F^{\dagger}(t)=
 e^{(L_0+\lambda \mathcal Q L_1)\,t}F+ O(\lambda\,\delta^2).
\label{O}
\end{eqnarray}
In what follows we shall retain in the expansion of $F^\dagger(t)$ only terms
up to second order in $\lambda$, 
\begin{eqnarray}
F^{\dagger}(t)\approx F_0(t)+\lambda\,F_1(t)+\lambda^2\, F_2(t).
\label{expansionB}
\end{eqnarray} 
The term $O(\lambda\,\delta^2)$ in (\ref{O}) does not contribute
to this approximation, because
$\lambda\,\delta^2=\lambda^3\epsilon^2$ is of order $\lambda^3$ for regime $B$
($\epsilon\sim 1$), or less for regime $A$ ($\epsilon\ll 1$). 
Applying the identity (\ref{aux2}) 
to the operator $\exp[(L_0+\lambda \mathcal Q L_1)\,t]$ in a recurrent manner, 
one obtains
\begin{eqnarray}
F_0(t)&=&e^{L_0t}F,\label{F0}\\
F_1(t)&=&\int_0^t d\tau e^{L_0(t-\tau)} \mathcal Q L_1 F_0(\tau),\label{F1}\\
F_2(t)&=&\int_0^t d\tau e^{L_0(t-\tau)} \mathcal Q L_1 F_1(\tau).\label{F2}
\end{eqnarray}
The term $F_0(t)$ is the pressure force, i.e., the force exerted by the bath on
the fixed particle. Terms $F_1(t)$ and $F_2(t)$ have no direct 
physical meaning and depend on the particle momentum. 
This dependence is to be explicitly extracted.

To the lowest order, one substitutes  $F^{\dagger}(t)\approx
F_0(t)$ into the pre-Langevin equation of motion (\ref{exacteqAB}) and retains
terms up to order $\lambda^2$. The relativistic 
nonlinear term cubic in $P_*$ is of order 
$\lambda^2\delta^2=\lambda^4\epsilon^2\lesssim \lambda^4$, and does 
not show up
in this approximation. As a result, one obtains the linear generalized
Langevin equation
\begin{eqnarray}
\frac{d}{dt}P_*(t)=\lambda\,F_0(t)-
\frac{\lambda^2}{p_T^2}\int_0^t d\tau\, P_*(\tau)\, C_0(t-\tau)
\label{LE2}
\end{eqnarray}  
with the memory kernel 
\begin{eqnarray}
C_0(t)=\langle F F_0(t)\rangle.
\label{C_0}
\end{eqnarray}
As discussed in the Introduction, the quasirelativistic description 
is expected to be asymptotically valid only in the limit of instantaneous 
point interactions. Therefore, the above equation must be
taken in the Markovian limit
\begin{eqnarray}
\frac{d}{dt}P_*(t)=\lambda\,F_0(t)-
\lambda^2\alpha_0\, P_*(t),
\label{LE3}
\end{eqnarray}  
with 
\begin{eqnarray}
\alpha_0=\frac{1}{p_T^2}\int_0^\infty C_0(t)\, dt=\frac{\beta}{m}
\,\int_0^\infty \!\!\langle F\,F_0(t)\rangle\, dt.
\end{eqnarray}
The equation for the true momentum $P=\lambda^{-1}P_*$ reads
\begin{eqnarray}
\frac{d}{dt}P(t)=F_0(t)-\gamma_0\, P(t),
\label{LE4}
\end{eqnarray} 
with 
\begin{eqnarray}
\gamma_0=\frac{\beta\, D_0}{M}, \qquad D_0=\int_0^\infty 
\langle F\,F_0(t)\rangle \,dt.
\label{fdt4}
\end{eqnarray}
Thus, for regimes $A$ and $B$ in the lowest order  in $\lambda$ one obtains
the Langevin equation of the same form as for the nonrelativistic
domain with the standard fluctuation-dissipation relation. 
Although the fluctuating  force $F_0(t)$ is governed
by the relativistic Liouville operator $L_0$ for the bath, 
this only modifies the values  of the damping coefficient $\gamma_0$
and the effective strength of the noise $D_0$. 
Otherwise, relaxation properties of the particle remain 
indistinguishable from that for the  nonrelativistic domain.

In order to take into account non-trivial relativistic effects,  we must
retain in the expansion for $F^{\dagger}(t)$ the higher-order terms.  Let us
adopt  the $\lambda^2$-order approximation (\ref{expansionB}) and  
evaluate the 
correlation $\langle F^{\dagger}(t)
F\rangle$ in the pre-Langevin equation  (\ref{exacteqAB}), extracting
explicitly the dependence on $P_*$. 
After some algebra the result can be presented in the  form
\begin{eqnarray}
&&\langle F F^{\dagger}(t)\rangle=
\langle F\, F_0(t)\rangle +\lambda^2 \langle F\, F_2(t)\rangle\nonumber\\
&&=C_0(t)
+\lambda^2\left[
\left(\frac{P_*}{m}\right)^2C_1(t)
+\frac{1}{m}\,C_2(t)
\right].
\label{kernelB}
\end{eqnarray}
Here   
$C_0(t)$ is the correlation function of the pressure force (\ref{C_0}),  
while functions $C_1(t)$ and $C_2(t)$ 
are expressed in terms of more complicated 
correlations
\begin{eqnarray}
\!\!\!\!\!\!\!\!\!
C_1(t)\!&=&\!\int_0^t \!\!\!dt_1\int_0^{t_1} \!\!\!dt_2
\,\,\langle\!\langle
G_0\,\, G_2(t, t_1, t_2)
\rangle\!\rangle,
\nonumber\\
\!\!\!\!\!\!\!\!\!
C_2(t)\!&=&\!\int_0^t \!\!\!dt_1\int_0^{t_1} \!\!\!dt_2
\,\,\langle\!\langle
G_0\, G_0(t\!-\!t_1)\,\,G_1(t, t_2)\rangle\!\rangle.
\label{C12}
\end{eqnarray}
Here we use the notations
\begin{eqnarray}
G_0(t)&=&F_0(t),\nonumber\\
G_1(t,t_1)&=&S(t\!-\!t_1)\,F_0(t_1), \nonumber\\
G_2(t, t_1, t_2)&=& S(t\!-\!t_1) \,S(t_1\!-\!t_2) \,F_0(t_2),
\end{eqnarray}
with the  operator $S(t)=e^{L_0t}\frac{\partial}{\partial X}$ and $G_0=G_0(0)$. 
The double angular brackets stands for cumulants
$\langle\!\langle A_1 A_2 \rangle\!\rangle=
\langle A_1\, A_2 \rangle-
\langle A_1\rangle\, \langle A_2 \rangle$ and 
$\langle\!\langle A_1 A_2 A_3\rangle\!\rangle=
\langle A_1 A_2 A_3 \rangle-
\langle A_1\rangle\langle A_2\rangle\langle A_3\rangle
-\langle A_1\rangle\langle\!\langle  A_2A_3\rangle\!\rangle-
\langle A_2\rangle\langle\!\langle  A_1A_3\rangle\!\rangle-
\langle A_3\rangle\langle\!\langle  A_1A_2\rangle\!\rangle
$.

Note that the result (\ref{kernelB}) for 
$\langle F^{\dagger}(t) F\rangle$ does not involve
a contribution of the first order in $\lambda$.
One can show that this contribution $\lambda\langle F F_1(t)\rangle$  
is proportional to
the correlation $\langle F \,G_1(t, t_1)\rangle$ which vanishes for
the homogeneous bath. 
Note also that expressions (\ref{kernelB}) and (\ref{C12})
are the same as the corresponding results for 
the nonrelativistic theory~\cite{PS}, except that
the bath dynamics propagator $L_0$ now is of the quasirelativistic 
form (\ref{Lbath}). Let us stress that functions $C_i(t)$ do not depend on
$P_*$, so the expression (\ref{kernelB}) presents the explicit dependence
of the  kernel $\langle F\,F^{\dagger}(t)\rangle$ on $P_*$ to order $\lambda^2$.
%This is because, 
%as can be seen from (\ref{O}), the specifically relativistic contribution with
%$L_3$ is of order $\lambda^3$ and does not
%show up in (\ref{kernelB}), which is evaluated to the second 
%order in $\lambda$,
%as required by Eq.(\ref{exacteqAB}). 

Substitution of (\ref{kernelB}) into the pre-Langevin equation
(\ref{exacteqAB}) and retaining terms up to order $\lambda^4$ 
(neglecting terms of order $\lambda^4\delta^2$) 
produces the generalized (non-Markovian) nonlinear
Langevin equation 
\begin{eqnarray}
\frac{d}{dt} P_*(t)&=&\lambda\, F^{\dagger}(t)
-\lambda^2\int_0^t \!\!d\tau\, M_1(\tau)\, P_*(t\!-\!\tau)\nonumber\\
&-&\lambda^4\int_0^t \!\!d\tau\, M_2(\tau)\,\, 
P_*^3(t\!-\!\tau)
\label{NMLE}
\end{eqnarray}
with  the memory kernels 
\begin{eqnarray}
M_1(t)&=&\frac{1}{p_T^2}\, C_0(t)-\frac{2\lambda^2}{m^2}\, C_1(t)+
\frac{\lambda^2}{m p_T^2}\, C_2(t), \nonumber\\
M_2(t)&=&\frac{1}{m^2p_T^2}\, C_1(t)-\frac{\epsilon^2}{2\,p_T^4}\, C_0(t),
\label{kernels}
\end{eqnarray}
where  correlations $C_i(t)$ are given by (\ref{C_0}) and (\ref{C12}).

As discussed above,  there is  no reason to believe that 
the quasi-relativistic approach is
satisfactory for any systems but with short-range binary collisions.
In such cases  memory effects  are negligible and one can apply  
the Markovian ansatz
\begin{eqnarray}
M_i(t)\,\to\, \delta(t)\, \alpha_i, \qquad \alpha_i=\int_0^\infty M_i(t)\, dt. 
\end{eqnarray}
This puts the above generalized Langevin equation into the local form
\begin{eqnarray}
\frac{d}{dt} P_*(t)=\lambda\, F^{\dagger}(t)
-\lambda^2\alpha_1\, P_*(t)
-\lambda^4\alpha_{2}\,\, P_*^3(t),
\label{MLE}
\end{eqnarray}
with the damping coefficients
\begin{eqnarray}
\alpha_1&=&\frac{1}{p_T^2}\,D_0-
\frac{2\lambda^2}{m^2}\,D_1+\frac{\lambda^2}{m\,p_T^2}\, D_2,
\label{alpha1}\\
\alpha_2&=&\frac{1}{m^2p_T^2}\,D_1-
\frac{\epsilon^2}{2p_T^4}\,D_0,
\label{alpha2}
\end{eqnarray}
where $p_T=\sqrt{m/\beta}$ and
\begin{eqnarray}
D_i=\int_0^\infty C_i(t)\, dt,\qquad i=0,1,2.
\end{eqnarray}

Compared to the $\lambda^2$-order Langevin equation (\ref{LE3}), two new
features appear in  Eq. (\ref{MLE}) of order $\lambda^4$. First,
as one can see from (\ref{alpha1}),
there are $\lambda^2$-order corrections to the linear damping coefficient 
$\alpha_0=p_T^{-2}\, D_0$.
These corrections do not involve the relativistic parameter $\epsilon$, and
therefore are purely classical.   
Second, and more interesting, a nonlinear dissipation term
emerges with the damping coefficient 
$\alpha_2$ given by (\ref{alpha2}).
The first term on the right-hand side of Eq. (\ref{alpha2})
is classical and   the second one is relativistic.

Note that  the nonlinear classical and relativistic contributions in 
Eq. (\ref{MLE})
are  of order $\lambda^4$ and $\lambda^4\epsilon^2$, respectively.
Therefore, this equation is perturbatively consistent in general
only for regime $B$ when $\epsilon\sim 1$. For regime $A$ ($\epsilon\ll 1$)
the $\lambda^2$-order approximation (\ref{expansionB}) for $F^{\dagger}(t)$
may be insufficient. For instance, if $\epsilon \sim \lambda$ then 
relativistic nonlinear corrections are of order $\lambda^6$. This would 
require the expansion of $F^{\dagger}(t)$ up to order $\lambda^4$ and dealing
with more complicated correlation functions.

Recall that Eq. (\ref{MLE}) is for the scaled momentum $P_*=\lambda\, P$.
The Langevin equation for the particle's true momentum  $P$ reads
\begin{eqnarray}
\frac{d}{dt} P(t)=F^{\dagger}(t)
-\gamma_1\, P(t)
-\gamma_2\, P^3(t).
\label{MLE2}
\end{eqnarray}
with damping coefficients
\begin{eqnarray}
\gamma_1&=&\lambda^2\alpha_1=\frac{\beta}{M}\, D_0-\frac{2}{M^2}\, D_1+
\frac{\beta}{M^2}\, D_2,\nonumber
\\
\gamma_2&=&\lambda^6\alpha_2=\frac{\beta}{M^3}\,D_1-\frac{\beta}{2M^3c^2}D_0.
\label{gammas}
\end{eqnarray}
Comparing these results with phenomenological fluctuation-dissipation
relations (\ref{gamma12}), one observes that the latter are recovered 
if $D_0$ is identified as the total noise strength $D$, while $D_1$ and $D_2$ 
both vanish or negligible,
\begin{eqnarray}
D_0\to D, \quad D_1\to 0,\quad D_2\to 0.
\end{eqnarray} 
Needless to say, neither of these conditions is satisfied in general. 

A  qualitatively new feature is
the presence of the new term involving $D_1=\int_0^\infty C_1(t)\, dt$ 
in the expression for the
nonlinear damping coefficient $\gamma_2$. 
As a result, the sign of $\gamma_2$ is not necessarily negative, as in the
phenomenological theory, but
depends on relative values of $D_0$ and $D_1$ and therefore on temperature. 
For a classical model it was found that $D_1/D_0=m\beta/6$ ~\cite{PS}. 
Using this as a rough estimation, 
%(nothing better is available at present)
one would get from (\ref{gammas}) or
(\ref{alpha2}) the expression
\begin{eqnarray}
\gamma_2=\frac{D_0}{2}\,\left(\frac{\beta}{m}\right)^2\,
\left(\frac{1}{3}-\epsilon^2\right),
\end{eqnarray} 
which is positive for regime $A$, $\epsilon\ll 1$, and also for a
sub-domain $\epsilon<1/\sqrt{3}$ of regime $B$.
%No results are currently available about the relation  between $D_0$ and $D_1$
%for the relativistic domain. 

\section{Langevin equation: regime  $C$}
One can show that 
the Langevin equation and fluctuation-dissipation relations
derived in the previous section retain their forms for regime $C$ also.  
The relevant small parameter now is 
$\delta=\lambda\,\epsilon$, the Liouville operator is given by (\ref{LC}),
$L=L_0+\delta\cdot L_1+\delta^3\cdot L_2$,
and the pre-Langevin equation has the form (\ref{exacteqC}) with 
$p_T=1/c\beta$.  Otherwise the derivation is similar to that for 
regimes $A$ and  $B$.

Substitution of the lowest-order approximation for the projected force $F^{\dagger}(t)\approx
F_0(t)$ into the pre-Langevin equation of motion (\ref{exacteqC})
yields, in the Markovian limit, the linear Langevin equation and
the fluctuation-dissipation relation, both in standard 
forms (\ref{LE4}) and (\ref{fdt4}). As for regimes $A$ and $B$, 
no relativistic effects 
show up in this lowest approximation except for the modified value of 
the damping parameter $\gamma_0$.

The higher-order approximation corresponds to the expansion
\begin{eqnarray}
F^{\dagger}(t)\approx F_0(t)+\delta\cdot F_1(t)+\delta^2\cdot F_2(t)
\label{FC}
\end{eqnarray}
with $F_i(t)$ still given by expressions (\ref{F0})-(\ref{F2}), 
but now with the operator $L_1$ defined by (\ref{L1C}).
As we already noted, the results for regime $C$ 
can be obtained  from those for 
regimes $A$ and $B$ by making the substitution (\ref{transition}), 
$\lambda\to\delta$ and $m\to p_T/c$. In particular,
for the correlation $\langle F\,F^{\dagger}(t)\rangle$, instead of
(\ref{kernelB}) one obtains 
\begin{eqnarray}
\!\!\!\!\!\!
\langle F F^{\dagger}(t)\rangle\!=\! C_0(t)
+\delta^2\!\left[
\left(\frac{c\,P_*}{p_T}\right)^2\!C_1(t)
+\frac{c}{p_T}\,C_2(t)
\right],
\label{kernelC}
\end{eqnarray}
with the same  functions $C_i(t)$. Substitution of this  
into the pre-Langevin equation 
(\ref{exacteqC}) and taking the Markovian limit leads
to the nonlinear Langevin equation for the scaled momentum 
\begin{eqnarray}
\frac{d}{dt} P_*(t)=\lambda\, F^{\dagger}(t)
-\delta^2\cdot\alpha_1\, P_*(t)
-\delta^4\cdot\alpha_{2}\,\, P_*^3(t)
\label{MLE_C}
\end{eqnarray}
with 
\begin{eqnarray}
\alpha_1&=&\frac{1}{p_T^2}\,D_0-
2\left(\frac{c\,\delta}{p_T}\right)^2\,D_1+\frac{c\,\delta^2}{p_T^3}\, D_2,
\label{alpha1C}\\
\alpha_2&=&\frac{c^2}{p_T^4}\,D_1-
\frac{1}{2p_T^4}\,D_0,
\label{alpha2C}
\end{eqnarray}
and $p_T=1/c\,\beta$. 
Then, as is easy to check, 
the equation for the true momentum $P=\lambda^{-1}P_*$ has the same form
(\ref{MLE2}) as for regime $B$ with the same fluctuation-dissipation relations
(\ref{gammas}).

\section{Moments and Thermalization}
Although the nonlinear Langevin equation (\ref{MLE2}) cannot be integrated in
an analytical form, the presented method is convenient
to describe 
relaxation processes perturbatively. 
As the  first example consider the relaxation 
of the first moment $\langle P(t)\rangle$ for, say,
regime $B$.
From (\ref{MLE})  one gets
\begin{eqnarray}
\frac{d}{dt}\langle P_*(t)\rangle=-\lambda^2\alpha_1 \,\langle P_*(t)\rangle 
-\lambda^4\alpha_2\,
\langle P_*^3(t)\rangle.
\label{q1}
\end{eqnarray}
Since the third moment $\langle P_*^3(t)\rangle$ enters this equation
multiplied by $\lambda^4$, it is sufficient to describe its dynamics 
in the lowest order in $\lambda$, 
\begin{eqnarray}
\frac{d}{dt}\langle P^3_*(t)\rangle=
-\lambda^2\alpha_3 \,\langle P_*^3(t)\rangle 
+\lambda^2\alpha_4\,
\langle P_*(t)\rangle,
\label{q2}
\end{eqnarray}
where $\alpha_3=3\,D_0/p_T^2$ and $\alpha_4=6\, D_0$.
In the phenomenological theory this equation is derived from the linear
Langevin equation under the assumption of the Gaussian random 
force~\cite{Coffey}, but it also can be derived microscopically
without this assumption [see Eq.(\ref{Pnlow}) below].
The closed system (\ref{q1}) and (\ref{q2}) is 
perturbatively consistent and describes the relaxation of
$\langle P(t)\rangle$ with nonexponential corrections
of order $\lambda^4$.

In order to describe $\lambda^4$-order dynamics of higher moments
$\langle P_*^n(t)\rangle$, $n>1$,  without the assumption of Gaussian noise
one needs the Langevin equations for powers $P_*^n(t)$. 
The derivation of these equations, first discussed 
for the non-relativistic domain in~\cite{Albers} 
and recently in~\cite{P_th}, is a straightforward
generalization of the method described above. The equations for 
$\langle P_*^n(t)\rangle$ can be used, in particular, to prove  
the particle's thermalization  towards the Maxwell-J\"uttner
distribution $\rho_{MJ}(P)$ (\ref{MJ}) for which the equilibrium  
moments in one dimension 
are
\begin{eqnarray}
\!\!\!\!\!\!\!\!\!\!\!\!
\langle P^{2n}\rangle_{eq}&=&\int_{-\infty}^{\infty} \rho_{MJ}(P)\, P^{2n}\, dP\nonumber\\
&=&(2n-1)!!\,\left(\frac{M}{\beta}\right)^n\,
\frac{K_{1+n}(\delta^{-2})}{K_1(\delta^{-2})},
\label{Pn_e}
\end{eqnarray}
where $K_i(x)$ is the modified Bessel function of the second kind.
To the leading order in $\delta^2$ this  expression reads
\begin{eqnarray}
\!\!\!\!\!
\langle P^{2n}\rangle_{eq}\approx(2n-1)!!\,\left(\frac{M}{\beta}\right)^n\,
\left[1+\left(n+\frac{n^2}{2}\right)\,\delta^2\right].
\label{Pn_e_approx}
\end{eqnarray}
%and particular for the second moment
%\begin{eqnarray}
%\langle P^{2}\rangle_e\approx\frac{M}{\beta}\,
%\left(1+\frac{3}{2}\,\delta^2\right).\label{P2E}
%\langle
%P^{4}\rangle_e\approx3\,\left(\frac{M}{\beta}\right)^2
%\left(1+4\,\delta^2\right).\label{P4E}
%\end{eqnarray}
In what follows we derive the equations for the moments 
$\langle P^{n}(t)\rangle$ of a weakly relativistic particle ($\delta\ll 1$)
and show explicitly that they converge to the equilibrium values
(\ref{Pn_e_approx}).

We shall assume that the temperature
corresponds to regime $B$;  the consideration for regimes 
$A$ and $C$ is similar.
Starting with the exact equation of motion for the powers of the scaled 
momentum
\begin{eqnarray}
\frac{d}{dt}P_*^n(t)=e^{Lt}L\,P_*^n,
\end{eqnarray} 
and using the operator identity (\ref{aux1}) for the propagator $e^{Lt}$, 
one gets
\begin{eqnarray}
\frac{d}{dt}P_*^n(t)=\lambda \,R(t)+\lambda\int_0^t e^{L(t-\tau)} 
\mathcal P\, L\,R(\tau)
\end{eqnarray} 
with the zero-centered projected force 
\begin{eqnarray}
R(t)=\lambda^{-1}e^{\mathcal Q Lt}\,L\,P_*^n
=n \, e^{\mathcal Q Lt}F\,P_*^{n-1}.
\end{eqnarray}
Proceeding with steps similar to those in
Sec. IV, one obtains  the  
pre-Langevin equation
\begin{eqnarray}
&&\!\!\!\!\!\!\!\!\!\!\!\!\!
\frac{d}{dt}P_*^n(t)=\lambda\,R(t)+
\lambda^2\int_0^t d\tau\,e^{L(t-\tau)}
\nonumber\\
&&\!\!\!\!\!\!\!\!\!\!\!\!\!
\times
\left\{
-\frac{1}{p_T^2}P_*+\frac{\partial}{\partial P_*}
+\delta^2\,\frac{1}{2p_T^4}\,P_*^3
\right\}\,
 \langle F \,R(t) \rangle.\label{Pn_exact}
 \end{eqnarray} 
To the lowest order in $\lambda$ the nonlinear term in this equation should
be omitted and
the projected force is
 \begin{eqnarray}
R(t)\approx n\,F_0(t)\, P_*^{n-1}\equiv R_0(t),
\label{R0}
\end{eqnarray}
where, recall, $F_0(t)=e^{L_0t}F$.
Substitution of this approximation into
the above pre-Langevin equation and taking 
the Markovian limit yields the Langevin equations for $P_*^n$
\begin{eqnarray}
\frac{d}{dt}P_*^n(t)=\lambda R_0(t) +k_1\,P_*^{n-2}(t)+k_2\,P_*^n(t)
\end{eqnarray}
with coefficients
\begin{eqnarray}
k_1=\lambda^2 n\,(n-1) D_0\, \qquad k_2=-\lambda^2 n\, D_0\,p_T^{-2},
\end{eqnarray}
where $D_0=\int_0^\infty \langle F F_0(\tau)\rangle$ and  
$p_T=\sqrt{m/\beta}$.
As expected, no relativistic corrections appear to the lowest 
perturbation order. The moments are governed by the equation
\begin{eqnarray}
\frac{d}{dt}\langle P_*^n(t)\rangle=k_1\,\langle P_*^{n-2}(t)\rangle
+k_2\,\langle P_*^n(t)\rangle, 
\label{Pnlow}
 \end{eqnarray}
and relax in the long time limit
to the equilibrium Maxwellian values. 
In particular 
\begin{eqnarray}
\langle P_*^2(t)\rangle \to p_T^2, \quad
\langle P_*^4(t)\rangle \to 3\,p_T^4.
\label{Pnmaxwell}
\end{eqnarray}

%For regime $C$ one needs to replace $\lambda$ by 
%$\delta$ and to recall that for this regime $P_*=\delta\cdot P$ and 
%$p_T=1/c\beta$. 

Consider now the expansion of the projected force to the second order in 
$\lambda$,
\begin{eqnarray}
R(t)\approx R_0(t)+\lambda\, R_1(t)+\lambda^2\, R_2(t),
\end{eqnarray}
where $R_0(t)$ is given by (\ref{R0}) and 
\begin{eqnarray}
R_1(t)&=&\int_0^t e^{L_0(t-\tau)} \mathcal Q \,L_1\, R_0(\tau),\nonumber\\
R_2(t)&=&\int_0^t e^{L_0(t-\tau)} \mathcal Q \,L_1\, R_1(\tau).
\end{eqnarray}
We need to evaluate the explicit
dependence on $P_*$ of the  correlation 
\begin{eqnarray}
\!\!\!\!\!\!\!\!
\langle F R(t)\rangle= \langle F R_0(t)\rangle+
\lambda\,\langle F R_1(t)\rangle+
\lambda^2\langle F R_2(t)\rangle
\label{aux5}
\end{eqnarray}
in the pre-Langevin
equation (\ref{Pn_exact}).
According to (\ref{R0}), the first term  is of the form
\begin{eqnarray}
\langle F R_0(t)\rangle=c_0(t)\, P_*^{n-1}.
\label{aux6}
\end{eqnarray}
The explicit evaluation shows that
the second term vanishes identically $\langle F R_1(t)\rangle=0$ due to
symmetry and the third term can be written as  
\begin{eqnarray}
\langle F R_2(t)\rangle&=&c_1(t)\,P_*^{n+1}+c_2(t)\,P_*^{n-1}\nonumber\\
&+&(n-1)(n-2)\,c_3(t)\,P_*^{n-3}.
\label{aux7}
\end{eqnarray}
In these expressions $c_0(t)=n\langle F F_0(t)\rangle$ and the other 
functions $c_i(t)$
are expressed in terms of more complicated correlation functions and do not
depend on $P_*$.  
Remarkably, as shown below, 
neither the explicit form of functions $c_i(t)$ nor their possible 
relations are  needed to prove the convergence of the moments to 
the equilibrium values (\ref{Pn_e}). 
[In the last term of Eq. (\ref{aux7}) we extracted explicitly
the factors $(n-1)(n-2)$ to make it clear that this term vanishes
for the first and second moments.]

Substitution of (\ref{aux5})-(\ref{aux7}) into (\ref{Pn_exact}), 
applying the Markovian limit and taking the average leads
to the following equation for the moments to order $\lambda^4$  
\begin{eqnarray}
\!\!\!\!\!\!\!\!\!
\frac{d\langle P_*^n\rangle}{dt}\!=\!r_1\langle P_*^{n-2}\rangle
\!+\!
r_2\langle P_*^{n}\rangle\!+\!r_3\langle P_*^{n+2}\rangle
\!+\!r_4\langle P_*^{n-4}\rangle
\label{Pnhigher}
\end{eqnarray}
with coefficients
\begin{eqnarray}
r_1&=&\lambda^2\,(n-1)\,b_0+\lambda^4\,(n-1)\,b_2\nonumber\\
&-&\lambda^4\,p_T^{-2}(n-1)(n-2)\, b_3,\nonumber\\
r_2&=&-\lambda^2\,p_T^{-2}\, b_0+\lambda^4\,(n+1)\,b_1
-\lambda^4\,p_T^{-2}\,b_2,
\nonumber\\
r_3&=&\lambda^2\,\delta^2\,(2\,p_T)^{-4}\, b_0-\lambda^4\,p_T^{-2}\, b_1,
\nonumber\\
r_4&=&\lambda^4\,(n-3)(n-2)(n-1)\,b_3,
\label{rs}
\end{eqnarray}
where $b_i=\int_0^\infty c_i(t) \, dt$.
Compared to the $\lambda^2$-order Eq. (\ref{Pnlow}), 
Eq.(\ref{Pnhigher}) shows that  to order $\lambda^4$,
the moment $\langle P_*^n\rangle$ is coupled in general not only with  
$\langle P_*^{n-2}\rangle$, but also with  $\langle P_*^{n+2}\rangle$
and $\langle P_*^{n-4}\rangle$.  Note that in Eqs. (\ref{rs}) 
the only relativistic correction is 
the first term ($\sim \delta^2$) in the expression for $r_3$.

Let us focus on the equation for the second moment
\begin{eqnarray}
\frac{d\langle P_*^2(t)\rangle}{dt}=r_1
+r_2\,\langle P_*^{2}(t)\rangle+r_3\,\langle P_*^{4}(t)\rangle
\label{P2}
\end{eqnarray}
with
\begin{eqnarray}
r_1&=&\lambda^2\,b_0+\lambda^4\,b_2,
\nonumber\\
r_2&=&-\lambda^2\,p_T^{-2}\, b_0+3\lambda^4\,b_1-\lambda^4\,p_T^{-2}\,b_2,
\nonumber\\
r_3&=&\lambda^2\delta^2 \, (2\,p_T)^{-4}\,b_0-\lambda^4\,p_T^{-2}\, b_1.
\label{r_forP2}
\end{eqnarray}
Using  Laplace transformations
\begin{eqnarray}
A_n(s)=\int_0^\infty e^{-st}\langle P_*^n(t)\rangle\, dt,
\end{eqnarray}
the stationary  value of the second moment can be written as
\begin{eqnarray}
\lim_{t\to\infty}\langle P_*^2(t)\rangle&=&\lim_{s\to 0}s\,A_2(s)\nonumber\\
&=&
-\frac{1}{r_2}\,[\,r_1+r_3\, \lim_{s\to 0} s\, A_4(s)\,].
\label{aux8}
\end{eqnarray} 
The stationary value for the fourth moment 
$\lim_{s\to 0} s\, A_4(s)$ 
appears here multiplied by $r_3\sim \lambda^4$. 
Then, one should assign to it
the equilibrium value found above in the lowest perturbation order 
[Eq. (\ref{Pnmaxwell})]
\begin{eqnarray}
\langle P_*^4\rangle _{eq}=\lim_{s\to 0} s\, A_4(s)=3\,p_T^4.
\end{eqnarray}
Then from (\ref{aux8}) and (\ref{r_forP2}) one obtains
to order $\delta^2$
\begin{eqnarray}
\lim_{t\to\infty}\langle P_*^2(t)\rangle=
p_T^2\left(1+\frac{3}{2}\,\delta^2\right),
\end{eqnarray}
which is consistent  with the prediction (\ref{Pn_e_approx}) 
of the equilibrium theory
with the Maxwell-J\"uttner distribution.

Thermalization of the  
moments of higher orders can be considered in a similar way. In particular,
one can  show that for the fourth moment equation (\ref{Pnhigher}) leads to the 
asymptotic result
\begin{eqnarray}
\lim_{t\to\infty}\langle P_*^4(t)\rangle=
3\,p_T^4\left(1+4\,\delta^2\right),
\end{eqnarray}
which is the correct $\delta^2$-order approximation for 
the equilibrium value 
$\langle P_*^4\rangle_{eq}$ given by  Eq. (\ref{Pn_e_approx}).  

\section{Concluding remarks}
In this paper we argue that the conventional Langevin phenomenology,
with a single fluctuation-dissipation relation,
cannot be extended to the relativistic domain. 
For a non-relativistic Brownian particle
the Langevin equation can be recovered from microscopic dynamics
in the weak coupling limit, i.e., in the leading order in 
$\lambda$.
For a relativistic Brownian particle such a  procedure
is inconsistent because nonlinear relativistic 
corrections are of 
the same order of magnitude (or  even smaller, for regime $A$)  
as classical  corrections to the weak-coupling approximation. 
We believe that this conclusion is to  be valid in general, even
though the presented theory employs the quasirelativistic 
approximation.

The necessity to go beyond the weak-coupling limit leads to 
more than one and more complicated
fluctuation-dissipation relations (\ref{gammas}). 
One interesting consequence 
is that the damping coefficient $\gamma_2$ of the nonlinear dissipation 
term  in the Langevin equation (\ref{MLE2}) may 
change its sign with temperature. This may lead to qualitatively different
relaxation behavior for different temperature intervals.  
For example, consider the ensemble of Brownian particles
for which  the initial first moment $\langle P(0)\rangle$  is zero, 
but the third moment $\langle P^3(0)\rangle$ is not. Then it can be 
shown~\cite{PF} that for $t>0$ the average momentum of the
the ensemble is temporarily nonzero, and its direction  
is determined by the sign of the dissipation 
coefficient $\gamma_2$.

In contrast to phenomenological models, the presented approach does not 
assume  that the fluctuating force in the Langevin equation is a Gaussian 
process. Fluctuation-dissipation relations involves cumulants of orders 
higher than 2, which in general do not vanish. With a non-Gaussian noise
many conventional methods of the phenomenological theory, for instance the
evaluation of higher moments,  cannot be applied.
Yet the perturbational approach developed in this paper
provides a 
systematic method to solve the equations of stochastic dynamics
analytically to any given order 
of a relevant small parameter. As an example, we discussed in Sec. VII
the thermalization problem and showed that 
the moments $\langle P^n(t)\rangle$ 
relax towards
the equilibrium values prescribed by the Maxwell-J\"uttner distribution,
provided this distribution holds for particles of the bath. 
Previously, the validity of the 
Maxwell-J\"uttner distribution was questioned
in a number of papers~\cite{Dunkel_paper1,Horwitz,Kaniadakis,Lehmann}, 
but was supported
by numerical simulations~\cite{MJ_simulation}.

The presented procedure can also be applied    
to derive the Fokker-Planck equation
for the distribution function $f(P,t)$. As is well 
known~\cite{Kampen_paper,Plyukhin_FP}, 
beyond the weak-coupling limit this equation in general contains 
derivatives with respect
to $P$ of orders higher than two and therefore is not of the form
(\ref{FPE}) implied 
%in the phenomenological approach
in phenomenological models with Gaussian noise.

%The assumption of Gaussian 
%noise can be justified when  the Brownian particle interacts simultaneously
%with the large number of bath particles~\cite{PS}. This is definetly not the
%case for systems with local  contact interactions discussed in this paper.

The  quasirelativistic approach adopted in this paper 
treats systems with finite-range interactions only approximately 
and contains no parameter that would describe qualitatively the validity 
of this approximation. This obliges one to restrict the application to systems
with collision-like interactions, which can be described in the Markovian
limit. A systematic incorporation of non-Markovian effects 
requires a much more elaborate theory that would explicitly takes 
into account the fields' degrees of freedom.

As a final comment let us note that while the presented theory provides 
explicit microscopic expressions for the 
damping coefficients $\gamma_1$ and $\gamma_2$, it is not clear if there is
a general relation between the two quantities. Remarkably, 
%similar to the classical domain~\cite{P_th}, 
such a relation is not required to prove thermalization of the particle  
towards the Maxwel-J\"uttner equilibrium  distribution.

\acknowledgments
I appreciate discussions with G. Buck and J. Schnick.

%\end{multicols}

\end{document}